\documentclass[Physsubmission, Phys]{SciPost}

\usepackage[bitstream-charter]{mathdesign}
\DeclareSymbolFont{usualmathcal}{OMS}{cmsy}{m}{n}
\DeclareSymbolFontAlphabet{\mathcal}{usualmathcal}

\begin{document}

\begin{center}
{\Large \textbf{
Precision tau physics: Challenge for Theory, on and off the lattice
\\
}}
\end{center}

\begin{center}
Amarjit Soni\
\end{center}

\begin{center}
 {Physics Department, Brookhaven National Laboratory, Upton, NY 11973, USA}
* adlersoni@gmail.com
\end{center}

\begin{center}
\today
\end{center}


\definecolor{palegray}{gray}{0.95}
\begin{center}
\colorbox{palegray}{
  \begin{minipage}{0.95\textwidth}
    \begin{center}
    {\it  16th International Workshop on Tau Lepton Physics (TAU2021),}\\
    {\it September 27 – October 1, 2021} \\
    \doi{10.21468/SciPostPhysProc.?}\\
    \end{center}
  \end{minipage}
}
\end{center}

\section*{Abstract}
{\bf
$\tau$ is playing an important role in the current B-physics indications from experiments of lepton flavor universality violations(LFUV).
This suggests it be given increasing attention theoretically in the coming years, given also the fact that Belle-II will have much larger data samples to study; similar comments also apply to LHCb as well as ATLAS and CMS. The fact that $\tau$ decays provide valuable information on its spin is an added advantage. This gains special significance if the current indications of new physics are upheld since naturalness arguments then strongly suggest that new physics should be accompanied by new CP-odd phase(s). Moreover, the fact that the $\tau$ mass is around 1.8 GeV,~{\it i.e} a lot less than the B-meson mass makes it much easier candidate for lattice studies
in great detail and consequently very likely with greater precision. This should help us test the SM and or BSMs with greater precision as needed.

}

\vspace{10pt}
\noindent\rule{\textwidth}{1pt}
\tableofcontents\thispagestyle{fancy}
\noindent\rule{\textwidth}{1pt}
\vspace{10pt}

\section{Introduction and motivation}
For almost a decade now, there have been persistent indications from all three major B-facilities of lepton universality violations 
(LUV)~\cite{BaBar:2012obs}, ~\cite{Huschle:2015rga,Hirose:2016wfn,Abdesselam:2019dgh, Abdesselam:2019wac,Abdesselam:2019lab,Aaij:2015yra,Aaij:2017uff,Aaij:2017vbb, Aaij:2019wad}, with a combined significance of
$4.5\sigma$~\cite{Amhis:2019ckw}. 
In tree-level charge current decays, $B \to  D(*) \tau (l)  \nu$, the reactions involving $\tau$ seem to differ 
more from those involving muon or
electron than predicted by the SM. Another class of reaction that is presenting an even stronger indications of lepton 
flavor universality violations is flavor changing neutral current decays which, in the SM are loop-level. At the quark-lepton level 
this is $b \to s l^+ l^-$ for $l = \mu$ or e.
The corresponding ratios of such reactions, $R_{K(*)}$ etc which in the SM are supposed to be unity 
within O(few \%), experimentally at LHCb ~\cite{Aaij:2017vbb}  are found to be
significantly different from unity {\it i.e.}
O($\approx 30\%$) at around 3 or even possibly
$\approx 4 \sigma$.

Moreover, another rather strong indication of new physics (NP) that has been strengthened this year is the muon (g-2) anomaly. 
For well over a decade this was already around $3.5 \sigma$ being the difference between the experimental measurements at BNL from about 2002~\cite{ Muong-2:2002wip} to around       2006~\cite{Bennett:2006fi}
and the theory predictions based on dispersion relations, data driven approach~\cite{Jegerlehner:2009ry}. Recent confirmation by the Fermilab (g-2) Collaboration~\cite{Muong-2:2021vma}
of the old BNL result has now increased the deviation from the data-driven theory prediction from the SM  to about 4.2$\sigma$.

The main point to bear in mind is that there are three anomalies, $R_{D(*)}$, $R_{K(*)}$ and muon (g-2), and each of these is over three $\sigma$.

So, the chances of at least one of these surviving the test of time and leading us to new physics is rather high, possibly LUV new physics.
Then, natuarlness arguements strongly suggest CP-violating BSM-phase(s).
In the context of CP-violation searches, $\tau$ is especially important due to the fact 
that its decay products provide information on its polarization.

Belle-II with its promise of significantly improved luminosity in an inherently cleaner environment of $e^+-e^-$ collisions as well  as
LHC experiments ATLAS, CMS and LHCb with RUN2 and  later runs can all provide much larger data samples of $\tau$ with greater
precision which should be extremely valuable.

In so far as theory goes, while for many aspects of $\tau$ decays, $m_{\tau}$ is too large for chiral perturbation theory (ChPT),
and not sufficiently large for QCD perturbation theory to work too well, it is much better suited than say the B mass, for lattice
calculations for precise studies both for testing the SM as well as serving as a very importantly laboratory for precise studies at the B-mass.

\section{More on current anomalies}
 {\bf I.Two B-anomalies}
 
There are for now two rather serious (over 3$\sigma$ each) deviations from the SM.
\begin{itemize}

\item Charge current semi-leptonic tree-level B decays to pseudo-scalar (D)
or vector ($D^*$) exclusive hadronic final state accompanied by either $\tau + \nu_{\tau}$
or $l + \nu_l$ (for $l=\mu$ or e). To test lepton universality as predicted by the SM a simple ratio,
of rates is formed,

\begin{align}
 R_{D^(*)}^{\tau/\ell} = {\rm BR}(B \to D^{(*)} \tau \bar{\nu})/{\rm BR}(B \to D^{(*)} \ell \bar{\nu}) (with ~\ell=e,\mu)  
\end{align}

Completely analogously,
\begin{align}
R_{\psi}   =    B_c\to J/\psi\ \ell \nu ~ (with ~ \ell~=~\mu,~\tau)
\end{align}

\item  Then there are FCNC decays, at the quark level $b \to s$,  that occur in the SM at 1-loop order wherein comparison of muons to
electrons with exclusive hadronic final states, $B^+ \to K^+ (K^{+*})$,
\begin{align}
R_{K^{(*)}}^{\mu/e}={\rm BR}(B \to K^{(*)} \mu^+\mu^-)/{\rm BR}(B \to K^{(*)} e^+e^-)
\end{align}
show significant deviation over 3$\sigma$, O(30\%) from unity.  
\end{itemize}

It is important to note here that since experimentally dilepton invariant mass is restricted to be greater than $\approx$ 500 MeV,
the difference in radiative corrections between electrons and muons in the SM is  expected to be only O(few\%); this is negligible compared
to the difference in $R_{K^{(*)}}$ that LHCb finds.
It is clearly experimentally very important to extend this test to the $\tau^{+-}$ pair but the expected BR is too small for the currently available
luminosities at LHCb or at BELLE-II.

In so far as the ratios $R_{D^{(*)}}$ and $R_{\psi}$ are concerned they are completely independent of the mixing angle $V_{cb}$ but they do
require precise knowledge of the $q^2$ dependence of the relevant form factors although because they are ratios some dependence does get partly cancelled.
\vspace{0.2 in}

 {\bf II. muon (g-2) anomaly}  

For a long time, the anomalous magnetic moment of the muon has been a target
of search for physics beyond the SM~\cite{Roberts:2018vsx}. Modern era started
at Brookhaven National Lab about 30 years ago leading to BNL experiment E821.

E821 published two key results. In 2002 in a Phys. Rev. Lett. ~\cite{Muong-2:2002wip}
and there final result in 2006~\cite{Muong-2:2006rrc}.
These two experimental results although one was for $\mu^-$ and the other for $\mu^+$ were
completely consistent  with each other and in fact agreed to 6 decimal places.

Comparison of the experimental results with the data-driven theory prediction from
the SM, which uses well-grounded dispersion relation~\cite{Lautrup:1971yp,Czarnecki:2002nt, Czarnecki:2017rlm} showed that the BNL result of 2002 incompatible to about 2.7$\sigma$. Later in 2006 the final E821 result disagreed with the SM prediction  by about 2.2 to 2.7 $\sigma$~\cite{Muong-2:2006rrc}.

For the past eight or so years there has been a new major experimental effort, E989  at Fermilab with the express goal of improving the BNL result on muon (g-2) by a factor of four in accuracy.
Along with that experimental effort, there  has been also an extensive theory effort to improve the accuracy of predicted value of muon (g-2) based on the SM.

Early in 2021, the FermiLab (g-2) collaboration announced their 1st result~\cite{Muong-2:2021vma} which they found to be completely consistent with the old  BNL result, in fact the two results agree again to 6 decimal places. The weighted average of the two experimental results disagrees with the data driven, "ratio method"~\cite{Aoyama:2020ynm} by $\approx$ 4.2 $\sigma$. In our search for new physics this is the major new positive development. This is because the data driven ratio method is very reliable.

Indeed these two anomalies, $R_{K(*)}$ and muon (g-2) are very serious, plausible each one of them is already over 4$\sigma$. Fermilab (g-2) collaboration is scheduled to announce their result with the addition of Run~2 data in about a year and likewise we expect update from LHCb with more data and  including also more channels, {\it e.g} FCNC decays of b-baryons in a similar time-frame. So, one or both of these anomalies could be over 5 $\sigma$ in a year.

The conclusion thus far is that chances of at least one of these two signals of new physics being genuine are very high. That provides us rationale to think of theoretical interesting possible new physics scenarios.

\section{BSM possibilties}

 Taking the experimental hints for lepton flavor violations (LFV) seriously, many have offered lepto-quark solutions~\cite{Fajfer:2015ycq, Bauer:2015knc,Crivellin:2017zlb,Das:2016vkr,
 Angelescu:2021lln}.   However, it seems to us~\cite{PhysRevD.96.095010,PhysRevD.102.015031,BhupalDev:2021ipu} that it is very likely that when nature is taking such a dramatically different  turn from the past, it will also very likely use the opportunity to  address a very glaring and pressing problem in the SM, namely the extreme radiative instability of the Higgs to radiative corrections. The simplest and most elegant solution for this problem remains the one offered  by supersymmetry.  But, in the current context of LFV the more natural version of SUSY is in fact  RPV-SUSY. Recall that conventionally, R-parity conservation has to be imposed to prevent proton decay. As Raman 
 {\it et.~al} have recently emphasized, the more natural setting of supersymmetry is in fact RPV-SUSY~\cite{Brust:2011tb}.
 
 Indeed, in our~\cite{PhysRevD.96.095010} first paper on the subject, to address the experimental hint in $B \to D(*) \tau(e,\mu) \nu$, for simplicity and minimality we assumed that the 3rd generation superpartners were the lightest.To emphasize the crucial role of the 3rd generation, we termed our version as "RPV3-SUSY". We then explicitly showed in that 1st work that even with one relatively light generation of superpartners, one of the most attractive feature of SUSY,
 namely gauge coupling unification stays intact along with, of course, the radiative stability of the Higgs.
 
 Another point that we have never seen emphasized before is that in fact as an inherent consequence of Bose-Fermi symmetry of the RPV-SUSY, it follows from a simple   generalization of the Yang-Mills idea~\cite{Yang:1954ek} that all interactions allowed by the internal symmetry  leads naturally to lepton flavor violations and removes the accidental symmetry of individual lepton number conservation that is there in the SM.
 
 {\bf \it Notable constraints and key experimental consequences of RPV3-Susy}
 
 It took significant effort on our part~\cite{    PhysRevD.96.095010,PhysRevD.102.015031} to seriously study and impose dozens of constraints on the RPV3
 set up we had assumed. See, {\it e.g}, Table IV and figs.~6,~7~and~8 in ~\cite{PhysRevD.102.015031}
 
 A very important consequence of seriously imposing all the existing constraints is that the key super-partner relevant for B-LFV-anomalies, the sbottom has to lie in the range of $\approx$
 3 to about 12 TeV. In turn, this has the important consequence that the contribution of the sbottom to solution of the muon (g-2) anomaly gets severely limited. Indeed, in our RPV3 scenario, most of the contribution to the muon (g-2) anomaly arises from the exchange of a super-partner of $\nu_{\tau}$ which is relatively quite light, around 1 TeV. Not only this should be readily available at the LHC; in fact it leads to a spectacular signature of 4 charged muons!

 \section{New physics repercussions for the tau}  
 
 {\bf \it BSM-CP odd phase(s) and $\tau$ decays}
 
 As discussed in the preceding sections currently there are very strong indications of new physics from B-decays as well as from muon (g-2) anomalies.
 The key point is that BNL 1964 Nobel prize winning  experiment~\cite{Christenson:1964fg} of Cronin-Fitch
 et al demonstrated for the 1st time that CP is not a symmetry of nature.Thus naturalness arguments strongly suggest that new physics should be accompanied by new BSM-CP-odd phase(s).
 
 The point we want to emphasize is that $\tau$ decays are an extremely sensitive probe of CP violation effects. In part this is because decay products of the $\tau$ provide information on its spin. In part also because tau has an appreciable Br into multi-body final states, The reason is that once one has at least 4 linearly independent momenta (including spin)
 all CP violating effects whether they are odd under naive Time-Reversal ($T_N$-odd) or even under naive Time-Reversal become accessible~\cite{Atwood:2000tu}.

With at least 4-linearly four momenta, one can construct an anti-symmetrical tensor a CP-odd, $T_N$-odd observable. Then because of the CPT theorem one does not need a CP-even phase.
This has the advantage that such a CP violating observable can arise at tree-level. This is unlike the case of a CP-violating observable that is even under naive time reversal such as is case for an energy asymmetry or a partial rate asymmetry when one compares particle decays to a final state with the anti-particle decays to a conjugate  final state. Then a non-vanishing CP-violating asymmetry can only arise if there is also a CP-even phase. The presence of such  non-vanishing CP-even phase requires the Feynman amplitude be complex. This in turn requires the calculation be at least  at one-loop order, so that will carry a suppression factor of 
$\alpha_s/\pi$~\cite{Bander:1979px}. Thus, as a rule a tree-level CP violating effect may be appreciably larger.
However, there is another non-trivial requirement to manifestly demonstrate that this tree-level effect originating from an observable such as a triple product $T_N$-odd observable is genuinely CP-violating. This requires experimentally comparing  particle decays with an antiparticle to ensure that the observable is non-vanishing in both cases and in fact the asymmetries will have to be added~\cite{Atwood:2000tu}.

{\bf {\it Possible CP violation in $\tau^- \to \nu_{\tau}\pi^- K_S(\ge \pi^0)$}}

This is an extremely interesting final state that the BABAR Collaboration focused on to search for CP violation~\cite{BaBar:2011pij}. Having 3 particles in the final state has the advantage that, in principle, one can also search for an energy asymmetry between the decays of a particle versus an antiparticle. In their experiment BABAR chose to focus only on the partial rate asymmetry (PRA). They found a 2.8 $\sigma$ deviation from the predictions of the SM. As it is explicitly mentioned in the BABAR paper the final state may have an unknown number of $\pi^0$'s,  since there are testing for CP
asymmetry so long as decays of $\tau^-$
as well as those of $\tau^+$ include similar neutrals that go undetected and included, it remains a valid test of CP violation. However, since the final state is known to have at least a neutrino (through lepton number conservation),
a charged pion and a $K_S$, both relatively clearly and cleanly identifiable, therefore, it would be extremely useful to search for an energy asymmetry also in the same final state. 
This can be done with relative easy by focusing on the energy distribution of the $\pi^-$ in $\tau^-$ decay and comparing with the energy distribution of the $\pi^+$ in $\tau^+$ decays. 
It is useful to stress here two points. First an energy asymmetry can easily be larger in a certain localized region of energy and be smaller in another region than the rate asymmetry. In this context, it is very important to emphasize that the PRA are severely constrained by the CPT theorem. When you add up all the non-vanishing PRA's in different channels there have to cancellations among them so that the life-time of $\tau^+$ equals that of $\tau^-$. As already mentioned below there is no good reason to believe that energy asymmetry will not be energy-dependent as a rule and consequently cancellations will arise as you integrate over the energy-distribution. Second, because $\pi^0$ and such neutrals can escape detection in a rate asymmetry, an energy asymmetry is likely a much safer test of CP-asymmetry, especially given that the final state readily allows one to do that. BABAR Collaboration (and Belle and Belle-II) are strongly urged to do so with existing and with increasing data sets.

{\bf {\it Illustrative examples of multi-particle final states in $\tau$ decays}} 

Here are some examples that should be distinctive experimentally and have enough  number of momenta that all different types of asymmetries can be searched for: \\
$ \tau^+ \to \bar \nu + \pi^+ + \pi^+ + \pi^-$; to $  \bar \nu + \pi^+ + K_S + K^+ + K^- $; $ \bar \nu + \pi^+ + K_{S}  + \pi^+  + \pi^-$ ;$ \bar \nu + K(*)^+ \pi^+ + 
\pi^-$ etc. 

In this list energies of conjugate mesons or invariant mass of conjugate pair of mesons as well as PRA can be used but all three of this type will require non-vanishing CP-even "strong" phase and there is no good reason to think that those phases will be not there or will be too small. In fact since different (possibly overlapping) resonances are likely to feed the final state those strong phases should be sizeable. Therefore they should help in the search for SM-CP as well as any BSM CP-odd phase(s). In fact the current hints of new physics in LFV
and or in muon (g-2) anomaly should motivate searches for a BSM CP-odd phase. In $\tau$ decays final states  such as $K_S, \pi^{+-}, K^{+-}$
should allow construction of naive Time Reversal ($T_N$)-odd observables. If any of these are found to be non-vanishing then the decays of the conjugate $\tau$ into conjugate final states should also be studied. A genuine violation of CP-symmetry requires both sets to be non-vanishing and consistent with each other including the sign.  

{\bf{\it $\tau_{edm}$ may be enhanced: Should be a high priority at BELLE-II}}

Prominent BSM explanation of the current anomalies are  LQ models, RPV-SUSY or Z'. In all of these
cases the electric dipole moment of the $\tau$ can be significantly enhanced over the expectations from the SM. Direct measurements of the edm of the $\tau$ may be very challenging or simply not possible because of its very short life-time, {\it i.e.} $O(10^{-13})$ sec. Fortunately $\tau$ decays to multi-body
final states can be a very powerful probe. This should be a very high priority at Belle-II where $tau$ production and decays can be studied with relative easy.
A traditional method is to examine the production cross-section near threshold
to extract the dipole moment. But actually use of the multi-body final states is a much more sensitive probe of all types of CP asymmetries~\cite{Atwood:1991ka     }. From that information one can try to extract the value of the edm. In fact since a large number of possible observables are available for extracting edm in that
work, for the first time a simple theorem is proved to choose an "optimal observable", namely that will necessarily give  you best bound or the value that will have the minimum statistical error. Indeed, in recent years, this theorem forms the basis for machine learning techniques. (See {\it e.g} footnote 3 (Ref. 27) in~\cite{   Brehmer:2019xox      } or Ref. 43 in~\cite{     Brehmer:2019gmn           }).

{\bf {\it Signals of BSM may be (easier) in (CP-conserving) magnetic moment,  absolute rates,  CKM unitarity etc: importance of $m_{\tau}$}}

Since the SM-CKM-phase is not expected to cause any CP-asymmetry in $\tau$
decays except possibly the induced indrect CP violation originating from the 
the mixing of the two neutral kaons, any significant detection of sizeable 
asymmetry would be a strong sign of a BSM-CP-odd phase. Consequently it also serves as a very strong indicator of new physics as well. However, it does  depend crucially on the size of the BSM phase.  While there is no good reason for the new CP-odd phase to be 
too small, clearly it can easily order 0.3 or 0.1. Even such a difference as O(3) can cause years of delay in experimental detection in the case it is about 0.1 compared to if it is 
around 0.3. 

Therefore, it is very important also to keep on searching for new physics via CP-conserving tests as well. One important example is the magnetic moment of the $\tau$; new physics can cause a significant enhancement over the SM. In an $e^+ + e^-$ machine, the magnetic moment of the $\tau$ can be readily searched along very similar lines as the edm search discussed in the preceding paras, namely production via $e^+ + e^-$ machine leading to weak decays into multi-particle final states. Again a multitude of observables can be used amongst which an "optimal observable" can be constructed following~\cite{      Atwood:1991ka              }.

Precise comparison of $\tau$ decays of  states containing muons versus those containing electrons can be a very valuable search for new physics especially lepton universality tests. 

In the class of CP-conserving tests, testing the SM via precise predictions of the BR against the experimental measured value can be immensely useful. To facilitate such tests lattice calculation of the partial widths can play a significant role. For such purposes the fact that mass of the  $\tau$ is only about 1.8 GeV, {\i.e.} it is substantially less than the mass of the b-quark is very significant. This is because lattice frameworks that preserve essentially the full chiral symmetry of the
continuum theory, such as "domain wall quarks" (DWQ), and as a result are more continuum-like (thus needing simpler renormalization as well as a simpler effective chiral Lagrangian for extrapolation purposes)  tend to be more computationally expensive and for them as well as many other lattice formulations the b-quark of mass ~4.5 GeV is computationally too demanding
and they have to resort to approximations. This is no longer the case for
the $\tau$ as simulations with lattice cut-offs of around 2 GeV are becoming quite common even  with DWQ both because of increased computational resources as well as
technical and algorithmic advances.

For precisely extracting weak mixing angles and testing CKM unitarity new promising ideas have been proposed for using inclusive B-decays. This could be enormously usefulin B-physics. For one thing there are long-standing few $\sigma$ tensions between inclusive (which are based on non-lattice continuum methods) and exclusive (a lot of these come from use of lattice techniques) extractions.
Moreover, if one could use inclusive modes then because the BRs are a lot bigger then that would make the test for CKM unitarity and or the search for new physics become as much more sensitive.

Using the $\tau$ as a testing ground for these new ideas has at least two significant advantages over the B-decays. First as already mentioned use of $\tau$ does not require an approximation in so far as the lattice cut-off goes.
In addition, for the $\tau$
already a huge amount of data exists which is not the case especially for the suppressed b $\to$ u transitions. For using  $\tau$ decays inclusively one can subdivide as those going with $V_{ud} = \cos_{\theta}$ and those going with $V_{us} = \sin_{\theta}$, where $\theta$ is the Cabibbo angle. Experimentally,  they can be distinguished obviously as those events with 0 strangeness (s) and those with $s=+~or~-1$.  

\section{Summary \& Conclusion}

Experiments at major B-facilities studying B-decays for almost a decade have been providing hints of lepton universality violations. Relative strong indications (over $3\sigma$) have persisted for almost two decades also in muon (g-2) measurements. 
These original BNL results have now been confirmed by measurements done in an entirely new, major experiments effort at Fermilab increasing the tension with the Standard Model to over 4$\sigma$. These experiments strongly suggest that $\tau$ lepton is being influenced by new physics. Then naturalness reasoning strongly suggests that the new physics will be accompanied by BSM-CP-odd phase(s). Spin of the $\tau$ and its decays into multi-particle final states as well as its electric dipole moment should be vigorously studied. 
CP-conserving tests such as rate measurements and magnetic moment measurements will also be very timely. BELLE-II with its anticipated increase in luminosity can especially be very valuable. But of course LHCb  as well as ATLAS and CMS can also play a very important role and continual increase in luminosity  are strongly urged. The fact that the mass of the $\tau$ is a lot less than the b-quark mass can be very useful both for testing the Standard Model as well as valuable laboratory for testing new theoretical ideas in non-perturbative lattice  efforts.

\section{Acknowledgements}
I wish to thank the organizers and in particular Emilie Passemar for the excellent organization and for the equally good program of Tau2021. Also i must thank Bhupal Dev for discussions and other help. This research was supported in part by USDOE
Contract \# DE-SC0012704.


\bibliography{ref.bib}

\nolinenumbers

\end{document}